\documentclass[a4paper,12pt]{article}

\usepackage{graphicx}
\usepackage{amsmath,amssymb}
\usepackage{natbib,setspace}

\def\a{\mbox{\boldmath$a$}}
\def\b{\mbox{\boldmath$b$}}
\def\e{\mbox{\boldmath$e$}}
\def\f{\mbox{\boldmath$f$}}
\def\g{\mbox{\boldmath$g$}}
\def\h{\mbox{\boldmath$h$}}
\def\m{\mbox{\boldmath$m$}}
\def\q{\mbox{\boldmath$q$}}
\def\v{\mbox{\boldmath$v$}}
\def\u{\mbox{\boldmath$u$}}
\def\w{\mbox{\boldmath$w$}}
\def\y{\mbox{\boldmath$y$}}
\def\B{\mbox{\boldmath$B$}}
\def\D{\mbox{\boldmath$D$}}
\def\I{\mbox{\boldmath$I$}}
\def\K{\mbox{\boldmath$K$}}

\def\R{\mbox{\boldmath$R$}}
\def\S{\mbox{\boldmath$S$}}
\def\V{\mbox{\boldmath$V$}}
\def\W{\mbox{\boldmath$W$}}
\def\balpha{\mbox{\boldmath$\alpha$}}
\def\bbeta{\mbox{\boldmath$\beta$}}
\def\bomega{\mbox{\boldmath$\omega$}}
\def\btheta{\mbox{\boldmath$\theta$}}
\def\bfeta{\mbox{\boldmath$\eta$}}
\def\beps{\mbox{\boldmath$\varepsilon$}}
\def\bmu{\mbox{\boldmath$\mu$}}
\def\bPhi{\mbox{\boldmath$\Phi$}}
\def\bLambda{\mbox{\boldmath$\Lambda$}}
\def\bPsi{\mbox{\boldmath$\Psi$}}
\def\bOmega{\mbox{\boldmath$\Omega$}}
\def\bSigma{\mbox{\boldmath$\Sigma$}}
\def\zero{\mbox{\boldmath$0$}}
\def\one{\mbox{\boldmath$1$}}
\def\diag{\mathrm{diag}}

\def\aa{{\bf *}}
\def\ss{\hspace*{.45ex}}

\begin{document}

\title
{Skew selection for factor stochastic volatility models}

\author
{\Large{Jouchi Nakajima}%
\footnote{
Bank for International Settlements. E-mail: jouchi.nakajima@bis.org. The views expressed herein are those of the author alone and do not necessarily reflect those of Bank for International Settlements.}
\\[12mm]
January 2019
}

\date{}

\maketitle

\begin{abstract}
This paper proposes factor stochastic volatility models with skew error distributions. The generalized hyperbolic skew t-distribution is employed for common-factor processes and idiosyncratic shocks. Using a Bayesian sparsity modeling strategy for the skewness parameter provides a parsimonious skew structure for possibly high-dimensional stochastic volatility models. Analyses of daily stock returns are provided. Empirical results show that the skewness is important for common-factor processes but less for idiosyncratic shocks. The sparse skew structure improves prediction and portfolio performance.
\\


\end{abstract}

%%%%%%%%%%%%%%%%%%%%%%%%%%

\newpage

\section{Introduction}

Multivariate analyses of stochastic volatility (SV) have been widely discussed in both the theoretical and empirical contexts of financial econometrics literature~\citep[see e.g.,][] {ChibOmoriAsai09}. Among them, the class of factor stochastic volatility (FSV) models is a theoretically and practically effective approach to understanding dynamic variances and covariances in multivariate series of financial variables~\citep{GewekeZhou96, Aguilar1999, PittShephard99b, AguilarWest00, Han05, ChibNardariShephard06}. Summarizing common dynamics among the series by latent factors yields an efficient modeling strategy for multivariate volatility, and for high-dimensional applications in particular. Previous studies have proposed generalizations on the basis of a standard FSV model, for example, with time-varying factor loadings~\citep{LopesCarvalho07, CarvalhoLopesAguilar11, NakajimaWest12jfe, ZhouNakajimaWest14}, and heavy-tailed error distributions~\citep[e.g.,][]{ChibNardariShephard06, IshiharaOmori17, LiScharth18}.

The skewness for a return distribution of the financial variables has been recognized as an important aspect of model fit and predictive performance~\citep[e.g.][]{Hansen94, EberleinKellerPrause98, AasHaff06}. The skew error distributions have been incorporated into univariate SV models~\citep{NakajimaOmori12, AbantoValleLachosDey15, LangrockMichelotSohnKneib15, Kobayashi16, TakahashiWatanabeOmori16} and a Cholesky-type multivariate SV~\citep{Nakajima17}. Little is known, however, about the skew error distribution of the FSV models.

To fill this gap, this paper introduces an FSV model with skew error distribution. The generalized hyperbolic (GH) skew $t$-distribution~\citep[see e.g.,][]{Barndorff-NielsenShephard01,AasHaff06} is incorporated into common-factor processes and the idiosyncratic shocks in the FSV model. It is simple and straightforward to assume that each of the factor processes and idiosyncratic shocks is associated with a different skewness parameter. Given this full model, a key question is whether all of the skewness parameters are necessary to describe the skewness in the multivariate series of financial variables. Because the factor processes summarize the common dynamics in multivariate SV series, some of the skewness parameters in the idiosyncratic shocks may not be relevant, and furthermore, not all of the latent processes may need the skew distribution. If so, then, assuming all possible skewness parameters can lead to \textit{more} uncertainty in model-fit estimation and prediction, especially for high-dimensional applications.

To address this point, the paper proposes a skew selection strategy for the proposed FSV model that includes the skew error distribution. It employs Bayesian sparsity modeling, which has been widely exploited in the literature~\citep[e.g.,][]{GeorgeMcCulloch93, GeorgeMcCulloch97,  West2003,ClydeGeorge04}. A so-called spike-and-slab prior selects a zero or non-zero skewness parameter for each factor process and each idiosyncratic shock. This approach allows the proposed FSV model to select-out redundant skewness from the full model and to provide a parsimonious multivariate skew structure. The proposed approach can easily scale up, which has practical relevance for high-dimensional problems. The paper is related to the literature on Bayesian sparsity modeling in the context of multivariate SV for reducing the number of parameters. \cite{LoddoNiSun11} propose a Bayesian stochastic search approach; \cite{Kastner18} develops a global-local shrinkage prior; and \cite{NakajimaWest12jfe} and  \cite{ZhouNakajimaWest14} incorporate time-varying sparsity with a latent thresholding process.

This paper provides two empirical analyses using multivariate daily stock returns to illustrate the proposed modeling framework. The first analysis illustrates the key characteristics of the proposed model with a concise dataset of five US-sector indices. A forecasting exercise and a Value-at-Risk (VaR) analysis for portfolio are examined. In the second analysis, a larger-scale dataset of 20 US individual stock returns is analyzed, which highlights the advantages of the skew FSV model and the sparsity skewness modeling in a high-dimensional, practical setting. Portfolio performance in an out-of-sample forecasting exercise is provided. Empirical results from these analyses show that the skewness is important for common-factor processes but less for idiosyncratic shocks. The sparse skew structure contributes to improving stock returns and VaR forecasts as well as portfolio performance.

\cite{LiScharth18} develop the same type of the skew FSV model, incorporating the sparsity prior for the skewness parameter, as well as for the leverage-effect parameter. Their empirical analysis using daily stock returns shows that the FSV model with skewness performs better than a no-skew model for VaR forecasts. The focus of the current paper is different; it concentrates on how a sparse skew structure plays a role in improving the forecasting ability of the skew FSV model. The empirical analyses use several diagnostics of forecasting ability in addition to the VaR measure and reveal that the skew FSV model with the skew selection mechanism performs better than the one without it.

Section 2 formulates the FSV models that includes the skew distribution. Section 3 explains a Bayesian estimation strategy for model fitting. Section 4 provides the first empirical analysis using five US-sector stock indices. Section 5 shows the second analysis using 20 individual stock returns. Section 6 concludes.

%%%%%%%%%%%

\section{The model}

This section describes the GH skew $t$-distribution in a univariate case (Section \ref{sec:GH}), and then defines the FSV models with skew distribution (Section \ref{sec:FSV}). The identification issue for the factor loadings is addressed (Section \ref{sec:ident}), and the sparsity structure for the skewness is introduced (Section \ref{sec:sparse}).

%%%%

\subsection{GH skew $t$-distribution} \label{sec:GH}

The random variable that follows the GH skew $t$-distribution, defined by $w_{it}$ ($t=1,2,\ldots$) for a series $i$, can be written in the form of \textit{normal variance-mean mixture}:
	\begin{eqnarray}
	w_{it} \,=\, m_i + \beta_i z_{it} + \sqrt{z_{it}}\varepsilon_{it},
	\quad
	\varepsilon_{it} \sim N(0,1), \label{eq:w}
	\end{eqnarray}
where $z_{it}$ follows the generalized inverse gaussian (GIG) distribution~\citep[see e.g.,][]{Barndorff-NielsenShephard01,McNeil05, AasHaff06}. This original form of the GH skew $t$-distribution is quite flexible for determining a shape of distribution. Previous studies~\citep{Prause99, AasHaff06, NakajimaOmori12} discuss that the parameters in the GH skew-$t$ distribution are typically difficult to jointly estimate. Following \cite{NakajimaOmori12}, to derive a parsimonious structure of the skewed distribution for the SV models, the current paper assumes that (i) $z_{it} \sim IG(\nu_i/2,\nu_i/2)$, where $IG$ denotes the inverse gamma distribution; (ii) $m_i = -\beta_ic_i$, where $c_i \equiv \mathrm{E}(z_{it}) = \nu_i/(\nu_i-2)$, for $\mathrm{E}(w_{it}) = 0$; and (iii) $\nu_i>4$ for the finite variance of $w_{it}$. The resulting density function describes the shape of distribution with two parameters $\beta_i$ and $\nu_i$. Although the skewness and kurtosis of the distribution are functions of both these two parameters~\citep[see][]{AasHaff06}, the $\beta_i$ and $\nu_i$ essentially represent the degree of skewness and kurtosis, respectively.

The literature discusses several alternatives of skew $t$-distributions~\citep{Hansen94, FernandezSteel98, Prause99, JonesFaddy03, AzzaliniCapitanio03}. Specifications differ among the distributions, which yield different shapes of the density, tail behaviors in particular. This paper uses the GH skew $t$-distribution defined above to exploit the form of the normal variance-mean mixture, as it makes Bayesian computation for model-fitting easy and efficient as described later in the paper. While it is of interest to compare different specifications of the skew $t$-distributions in terms of model fit and forecasting, this paper sticks to the GH skew $t$-distribution because the main purpose of the analysis is to investigate the skew structure of financial time series in the context of FSV models. Running a horse race among the different skew $t$-distributions is left for future work.

%%%%

\subsection{Factor stochastic volatility with skew distribution} \label{sec:FSV}

Define a $k\times 1$ vector of response, $\y_t=(y_{1t},\ldots,y_{kt})'$, for $t=1,\ldots,T$. A standard FSV model is formulated as
	\begin{eqnarray}
	\y_t &=& \B \f_t + \e_t, \quad \e_t \sim N(\zero, \bLambda_t), \label{eq:obs} \\
	\f_t &\sim& N(\zero, \bPsi_t), \label{eq:factor}
	\end{eqnarray}
where $\f_t=(f_{1t},\ldots,f_{pt})'$ is a $p\times 1$ vector of the latent, common factors; $\B$ is a $k\times p$ matrix of factor loadings; $\bLambda_t = \diag(\exp(h_{1t}),\ldots,\exp(h_{kt}))$, and $\bPsi_t = \diag(\exp(h_{k+1,t}),\ldots,$ $\exp(h_{qt}))$, with $q=k+p$. This form of the model arises from the idea of dimension reduction in the unconditional variance of $\y_t$, denoted by $\bSigma_t$, as $\y_t\sim N(\zero, \bSigma_t)$. With a large $k$, the matrices in eqns.\ (\ref{eq:obs}) and (\ref{eq:factor}) yield an effective reduction of dimensionality in $\bSigma_t$, with $\bSigma_t=\B\bPsi_t\B'+\bLambda_t$. Because the matrices $\bPsi_t$ and $\bLambda_t$ are diagonal and a typical choice of $p$ is small, the number of parameters can be reduced effectively.

We build the proposed skew $t$-distribution on the simple form of the FSV model defined in eqns.\ (\ref{eq:obs}) and (\ref{eq:factor}). A variety of generalizations can be easily introduced, however. A local trend can be added to the observation equation (\ref{eq:obs}); the factor loading $\B$ can be time-varying; and the process of $\f_t$ can be dynamic, for example, as it follows a vector autoregressive (VAR) model~\citep[see e.g.,][] {ChibOmoriAsai09}. The following discussion on incorporating the skewness and its estimation method can be straightforwardly applied to the more elaborated FSV models.

The new class of the FSV models with the GH skew $t$-distribution is defined by
	\begin{eqnarray*}
	\y_t &=& \B \f_t + \bLambda_t^{1/2}\w_t, \quad \w_t = (w_{1t},\ldots w_{kt})', \\
	w_{it} &=& \beta_i (z_{it} - c_i) + \sqrt{z_{it}}\varepsilon_{it}, \quad i=1,\ldots,k, \\
	\f_t &=& \bPsi_t^{1/2}\v_t, \quad  \v_t = (v_{1t},\ldots v_{pt})', \\
	v_{jt} &=& \beta_{k+j} (z_{k+j,t} - c_{k+j}) + \sqrt{z_{k+j,t}}\varepsilon_{k+j,t}, \quad j=1,\ldots,p,
	\end{eqnarray*}
where each element of $\w_t$ and $\v_t$ follows the GH skew $t$-distribution specified by eqn.\ (\ref{eq:w}). Define $\h_t=(h_{1t},\ldots,h_{qt})'$, and $\beps_t=(\varepsilon_{1t},\ldots,\varepsilon_{qt})'$. The latent SV process is defined as the standard autoregressive form:
	\begin{eqnarray*}
	\h_{t+1} \,=\, \bmu + \bPhi(\h_t-\bmu) + \bfeta_t,
	\end{eqnarray*}
where $\bmu = (\mu_1,\ldots,\mu_q)'$, $\bPhi = \diag(\phi_1,\ldots,\phi_q)$, $\bfeta_t = (\eta_{1t},\ldots,\eta_{qt})$, and 
	\begin{eqnarray*}
	\left(\begin{array}{c} \beps_t \\ \bfeta_t \end{array}\right) \,\sim\, N\left( \zero, \left[ \begin{array}{cc} \I & \R \\ \R & \S \end{array}\right] \right),
	\end{eqnarray*}
where $\S = \diag(\sigma_1^2,\ldots,\sigma_q^2)$, and $\R = \diag(\rho_1\sigma_1,\ldots,\rho_q\sigma_q)$. This formulation makes each of the processes $\{h_{it}\}_{t=1}^T$ follow the univariate autoregressive process. The parameter $\rho$ measures the so-called leverage effect, which is widely addressed in financial econometrics~\citep[e.g.,][]{Black76, Nelson91, Yu05, OmoriChibShephardNakajima07}. In a stock market, for example, a large price drop tends to lead to a large increase in the volatility the next day. This behavior can be described using a negative correlation ($\rho<0$) between the two shocks $(\varepsilon_t, \eta_t)$.

A key aspect of the proposed model structure is that, conditional on $(\B, \f_t)$, the model reduces to $q$-fold state-space representation of the univariate SV model. To see this, define
	\begin{eqnarray*}
	\tilde{y}_{it} \,=\, \left\{ \begin{array}{ll} y_{it} - \B_{(i)}\f_t & (i \le k), \\ f_{jt} & (i>k;\,j=i-k), \end{array} \right. i=1,\ldots , q,
	\end{eqnarray*}
where $\B_{(i)}$ denotes a $1\times p$ vector of the $i$-th row of $\B$. Because of the diagonal structure of the matrices $(\bLambda_t, \bPsi_t, \S, \R)$, we obtain the univariate SV model with leverage effect, that is,
	\begin{eqnarray}
	\tilde{y}_{it} &=& \left\{ \beta_i (z_{it}-c_i) + \sqrt{z_{it}}\varepsilon_{it} \right\} \exp(h_{it}/2), \label{eq:sv1} \\
	h_{i,t+1} &=& \mu_i + \phi_i (h_{it} - \mu_i) + \eta_{it}, \\[1ex]
	\left( \begin{array}{c} \varepsilon_{it} \\ \eta_{it} \\ \end{array} \right)
	&\sim& N(\zero,\,\bOmega_i),
	\quad\mathrm{and}\quad
	\bOmega_i=
	\left( \begin{array}{cc} 1 & \rho_i\sigma_i \\ \rho_i\sigma_i & \sigma_i^2  \end{array} \right), \label{eq:sv3}
	\end{eqnarray}
for $i=1,\ldots,q$. In a Bayesian estimation framework, this structure enables us to easily scale up the dimension of the responses for the computation of parameter estimates. With the conditionally independent priors described below, we can obtain an efficient and fast parallel computation.

%%%%%%%%%%%

\subsection{Identification} \label{sec:ident}

A general modeling issue on the latent factors is restriction for identification. The factor models require a proper parameter restriction on the loading matrix or another valid approach. The most common approach, the one taken here, is the upper triangular form of the factor-loading matrix to order factors according to the selected first $p$ time-series variables~\citep[e.g.,][]{AguilarWest00, LopesWest04}. Specifically, we assume that $b_{ii}=1$, for $i=1,\ldots,p$; and $b_{ij}=0$, for $i<j$, and $i\le p$. Although this approach constrains the model because it depends on the selection and ordering of the first $p$ variables, it is practical to track it with a robustness check that examines different sets of the first $p$ series. Alternative strategies can be found in the existing studies~\citep[e.g.,][]{FruhwirthLopes18}.

%%%%%%%%%%%

\subsection{Sparsity prior for the skewness} \label{sec:sparse}

The Bayesian sparsity modeling strategy is incorporated into the skewness parameters to develop the skew selection in the FSV model. 
The sparsity prior for $\beta_i$ is given by
	\begin{eqnarray*}
	\beta_i \,\sim\, \kappa N(\beta_i|0, \tau_0^2) + (1-\kappa)\delta_0(\beta_i), \quad\quad i=1,\ldots,q,
	\end{eqnarray*}
where $\delta_0$ denotes the Dirac delta function at zero, and $0<\kappa<1$. This prior assigns the normal distribution for $\beta_i$ with probability $\kappa$ and the point mass at $\beta_i=0$ with probability $1-\kappa$. This is the so-called spike-and-slab prior that is often used in the literature~\citep[e.g.,][]{GeorgeMcCulloch93, GeorgeMcCulloch97,  West2003,ClydeGeorge04}. This approach yields an effective skew selection for the FSV model, selecting out redundant skewness parameters from the full set of skewness and providing parsimonious multivariate skew structure.

It is not easy to theoretically derive the skewness of $y_{it}$ in the resulting FSV model. The following brief simulation study addresses how the skewness of $y_{it}$ depends on the skewness parameters for the idiosyncratic shocks and the common-factor processes. A sample of size $T=1{,}000$, $k=3$, $p=2$ is simulated from the skew FSV model. The parameter values are set as $\phi_i=0.995$, $\sigma_i=0.05$, $\rho_i=-0.5$, and $\nu_i=8$ for $i=1,\ldots,q$. These values are selected following previous studies~\cite[e.g.][]{NakajimaOmori12, Nakajima17}. For the level of volatility, we set $\mu_i=-11$, for $i=1,2,3$, while $\mu_i=-10$ for $i=4,5$; to assume that the factor volatilities are higher than volatilities of idiosyncratic shocks, which are typically observed in practice. The free parameters in the factor loading matrix $\B$ are randomly generated from the uniform distribution $U[0.5,1.5]$. 

%%%%%%%%%%%%%%%%%%
%%% Parameters %%%
%%%%%%%%%%%%%%%%%%%%%%%%%%%%%%%%%% 
\begin{figure}[t]

{\small (i) $\bbeta=$\,(-1, -1, -1, 0, 0) \hspace{7mm} (ii) (0, 0, 0,  -1, 0) \hspace{6mm} (iii) (0, 0, 0, -1, -1) \hspace{5mm} (iv) (-1, -1, -1, -1, -1)} \\[-2mm]

\hspace{3mm}\includegraphics[scale = 0.97]{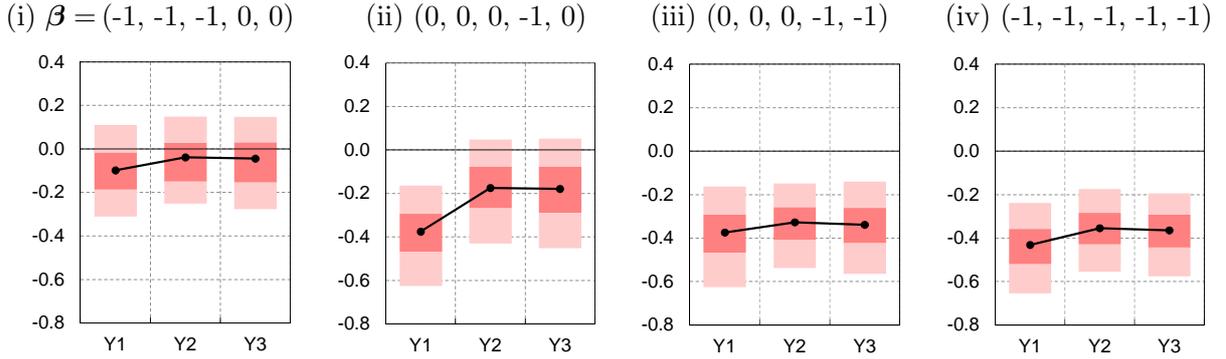}

\caption{Skewness of simulated data: medians (dot and solid line), the 50\% (filled area, dark) and 90\% (light) intervals for 1,000 sets of simulated series. The horizontal axis refers to the series (Y)}

\label{fig-sim}
\end{figure}
%%%%%%%%%%%%%%%%%%%%%%%%%%%%%%%%%%%%%%%%%%%%%%%%

Figure \ref{fig-sim} exhibits summaries of the skewness of $y_{it}$ from 1,000 sets of simulated data for different values of the skewness parameter: $\bbeta\equiv(\beta_1,\ldots,\beta_q)=$ (i) $(-1,-1,-1,0,0)$, (ii) $(0,0,0,-1,0)$, (iii) $(0,0,0,-1,-1)$, and (iv) $-\one_{1\times 5}$.  Case (i) assumes that only the idiosyncratic shocks follow the skew distribution; the resulting skewness of $\y_t$ is not so large as Case (iv) which assumes both the idiosyncratic shocks and the factor processes follow the skew distribution. This reflects the structure of the factor model: $y_{it}$ is the sum of the factor and the idiosyncratic shock. The level of the factor volatility is assumed to be higher than that of idiosyncratic shock in this simulation exercise, the symmetry of the factor processes weakens the degree of skewness caused from the idiosyncratic shock.

Case (ii) assumes that only the first factor process follows the skew distribution. The degree of skewness of the first series (Y1) is larger than that in Case (i), because the skewness in the first factor process dominates the idiosyncratic shock. In contrast, the degree of skewness in the second (Y2) and the third (Y3) series is slightly smaller than the first series, because the second factor, which follows the symmetric distribution but does not appear in the first series, weakens the degree of skewness. The level of volatility for the second factor, however, is typically smaller than that for the first factor in practice, which leads to a smaller difference between the skewness for the first series and the others. The key finding here is that the resulting skewness of $\y_{it}$ also depends on $\bmu$.

Case (iii) assumes that both two factor processes follow the skew distribution, which exhibits a notably similar degree of skewness as Case (iv). From this result, the assumption of $\beta_i\neq 0$ for all $i=1,\ldots,q$, as in Case (iv), can be redundant to describe the skewness of $\y_t$. The skew selection structure with the sparsity prior for $\bbeta$ is expected to effectively address this point for the skew FSV model, exploring the best parsimonious structure to select out the redundant parameters and improve the prediction performance.

%%%%%%%%%%%%%%%%%%%%%%%

\section{Bayesian estimation}

The Bayesian analysis and computation are developed for the proposed FSV model based on the Markov chain Monte Carlo (MCMC) method. The sampling estimation scheme can be built on popular efficient conditional samplers for univariate SV models with a leverage effect~\citep{OmoriChibShephardNakajima07, OmoriWatanabe08, NakajimaOmori12}, for the state space dynamic models~\citep[e.g.,][]{PradoWest10}, and for the multivariate FSV models~\citep[e.g.,][]{LopesCarvalho07, ChibOmoriAsai09, IshiharaOmori17}. \cite{LiScharth18} utilize particle Gibbs with an ancestor sampling algorithm to sample the volatility process in the skew FSV model. In contrast, the current paper shows that the full posterior analysis can be implemented by using the simpler algorithm of multi-move sampler for the volatility without any particle sampling.

Based on observations $\y = \{\y_1,\ldots,\y_T\}$, the conditional samplers are developed for each of the following latent variables and model parameters:
	\begin{enumerate}
	\item The $q$-fold univariate SV processes $\{h_{i1},\ldots,h_{iT}\}$, their parameters $\btheta_i \equiv \{\mu_i,\phi_i,\sigma_i,\rho_i,$ $\nu_i\}$, and mixing latent processes $\{z_{i1},\ldots,z_{iT}\}$ for $i=1,\ldots,q$.
	\item The factor process $\f_t$ and the loading matrix $\B$.
	\item The skewness parameter $\bbeta$ and the sparsity parameter $\kappa$.
	\end{enumerate}
A key computational strategy here is decoupling the univariate SV processes in the conditional samplers. Generally, a sampler for the SV process requires a relatively heavier computational burden and makes the mixing of the MCMC algorithm slower. Its simple multivariate extension and, further, with the skewness distribution can be more challenging. However, as mentioned before, the representation of the normal variance-mean mixture leads to the efficient computation with those conditional samplers for the SV processes and their parameters. In addition, the conditional independence of the univariate SV leads to a conditional sampler for $\beta_i$ under the sparsity prior.

Each of the conditional sampler is outlined as follows.

%%%%%%%%%%

\vspace{5mm} \noindent
\textbf{Generation of SV and mixing latent process} \\[1mm]
Based on eqns.\ (\ref{eq:sv1})--(\ref{eq:sv3}), each univariate SV process is generated using the conditional sampler for the SV with leverage effect. \cite{OmoriWatanabe08} develop the efficient multi-move sampler for this type of non-linear state space model and \cite{NakajimaOmori12} utilize it on the SV model with the GH skew $t$-distribution, which is taken in the current analysis. For the generation of the mixing latent process $z_{it}$, the conditional posterior distribution is the product of the kernel of the inverse gamma distribution and another part. A standard Metropolis-Hasting algorithm is employed with a candidate generated from the inverse gamma distribution. 

To develop the conditional sampler for the parameters defining each of the SV process, we simply assume traditional priors that are independent across $i$: normal priors for $\mu_i$, shifted beta priors for $\phi_i$ and $\rho_i$, inverse gamma priors for $\sigma_i^2$, and truncated gamma priors for $\nu_i$ with $\nu_i>4$. Then, the conditional samplers are implemented with standard posterior distributions for $\mu_i$ and the MH algorithm for ($\sigma_i,\rho_i$) and $\nu_i$~\citep[see][]{NakajimaOmori12}.

Note that because these latent processes $\{h_{i1},\ldots,h_{iT}, z_{i1},\ldots,z_{iT}\}$ and the parameter $\btheta_i$ are conditionally independent across $i$ in the conditional posterior inference, we can implement parallel computing for the generation of $q$-fold SV processes.

%%%%%%%%%%

\vspace{5mm} \noindent
\textbf{Generation of factor process and loadings} \\[1mm]
Define $\hat{\eta}_{it} = (h_{i,t+1} - \mu_i) - \phi_i(h_{it} - \mu_i)$, and $\hat{\sigma}_{it}^2 = z_{it}(1-\rho^2_i)e^{h_{it}}$, for $i=1,\ldots,q$. Conditional on the mixing latent process, SV process and its parameters, we obtain
	\begin{eqnarray*}
	y_{it} &=& \B_{(i)}\f_t + \alpha_{it} + \hat{u}_{it},\quad i=1,\ldots,k, \\
	f_{jt} &=& \alpha_{k+j,t} + \hat{u}_{k+j,t}, \quad j=1,\ldots,p,
	\end{eqnarray*}
where
	\begin{eqnarray*}
	\alpha_{\ell t} &=& \{\beta_i(z_{\ell t}-c_\ell ) + \sqrt{z_{\ell t}}\rho_\ell \hat{\eta}_{\ell t}/\sigma_i\}e^{h_{\ell t}/2}, \\
	\hat{u}_{\ell t} &\sim& N(0,\hat{\sigma}_{\ell t}^2),
	\end{eqnarray*}
for $\ell=1,\ldots,q$.

Define $\b_i=(B_{i1},\ldots,B_{ir})'$ as the $r\times 1$ vector of free parameters in the $i$-th row of $\B$, with $r=\mathrm{min}(i-1,p)$, for $i=2,\ldots,k$. We generate the sample of $\b_i$, for $i=2,\ldots,k$, based on the following regression model:
	\begin{eqnarray*}
	\hat{y}_{it} \,=\, d_{it} + \g_{it}'\b_i + \hat{u}_{it}, \quad \hat{u}_{it} \sim N(0,\hat{\sigma}_{it}^2),
	\end{eqnarray*}
where $\hat{y}_{it} = y_{it} - \alpha_{it}$, $\g_{it} = (f_{1t}, \ldots, f_{rt})'$, and
	\begin{eqnarray*}
	d_{it} \,=\, \left\{ \begin{array}{ll} f_{it} & (\mathrm{for}\ i=2,\ldots,p), \\
	0 & (\mathrm{for}\ i=p+1,\ldots,k). \end{array} \right. 
	\end{eqnarray*}
We set a prior as $\b_i \sim N(\b_{i0}, \W_{i0})$. Then, the conditional posterior distribution of $\b_i$ is $N(\hat{\b}_i, \hat{\W}_i)$, where
	\begin{eqnarray*}
	\hat{\W}_i \,=\, \left(\W_{i0}^{-1} + \sum_{t=1}^T \frac{\g_{it}\g_{it}'}{\hat{\sigma}_{it}^2} \right)^{-1},\quad 
	\hat{\b}_i \,=\, \hat{\W}_i \left(\W_{i0}^{-1}\b_{i0} + \sum_{t=1}^T \frac{(\hat{y}_{it} - d_{it})\g_{it}}{\hat{\sigma}_{it}^2} \right).
	\end{eqnarray*}

For the generation of $\f_t$, define $\hat{\y}_t = (\hat{y}_{1t},\ldots,\hat{y}_{kt})'$, and $\hat{\u}_t = (\hat{u}_{1t},\ldots,\hat{u}_{kt})'$. We obtain the following form:
	\begin{eqnarray*}
	\hat{\y}_t &=& \B \f_t + \hat{\u}_t, \\
	\f_t &\sim& N(\balpha_t, \hat{\bPsi}_t),
	\end{eqnarray*}
where $\balpha_t = (\alpha_{k+1,t},\ldots,\alpha_{qt})'$, and $\hat{\bPsi}_t = \diag(\hat{\sigma}_{k+1,t}^2,\ldots,\hat{\sigma}_{qt}^2)$. We generate $\f_t$ from the conditional posterior distribution, $N(\hat{\a}_t, \hat{\V}_t)$, where $\hat{\V}_t = (\hat{\bPsi}_t^{-1} + \B'\hat{\bSigma}^{-1}\B)^{-1}$, and $\hat{\a}_t = \hat{\V}_t (\hat{\bPsi}_t^{-1}\balpha_t + \B'\hat{\bSigma}_t^{-1}\hat{\y}_t)$, with $\hat{\bSigma}_t = \diag(\hat{\sigma}_{1t}^2,\ldots,\hat{\sigma}_{kt}^2)$, for $t=1,\ldots,T$.

%%%%%%%%%%

\vspace{5mm} \noindent
\textbf{Generation of skewness parameters} \\[1mm]
Before we derive the posterior distribution of the $\beta_i$ under the sparsity prior, consider a simple normal prior $N(0,\tau_0^2)$ for $\beta_i$. Then, the conditional posterior distribution of $\beta_i$ appears to be the normal distribution; we denote it as $N(\hat{\beta}_i,\hat{\tau}_i^2)$. Turning into the sparsity prior, the conditional posterior distribution is derived as conjugate as
	\begin{eqnarray*}
	\beta_i\,|\,\cdot \,\sim\, \hat{\kappa}_i N(\beta_i | \hat{\beta}_i,\hat{\tau}_i^2) + (1-\hat{\kappa}_i)\delta_0(\beta_i),
	\end{eqnarray*}
where $\hat{\kappa}_i=\kappa \gamma_i / (\kappa \gamma_i+1-\kappa)$, and $\gamma_i=\exp(\hat{\beta}_i^2/2\hat{\tau}_i^2)\hat{\tau}_i/\tau_0$. In this conditional posterior, $\beta_i$ follows the normal distribution with probability $\hat{\kappa}$, and shrunk at zero otherwise.

For the generation of $\kappa$, we set a beta prior and derive the conjugate posterior distribution conditional on the number of $\beta_i$'s such that $\beta_i \neq 0$.

%%%%%%%%%%%%%%%%%%%%%%%%%

\section{A study of US-sector indices}

The first real-data analysis applies the proposed FSV model to a series of $k=5$ daily stock returns of US-sector indices. This analysis demonstrates how the skew error distribution and sparsity modeling for the skewness parameter play the role of improving the model fit and forecasting ability of the FSV model.

%%%%%%%%%%%%%%%%%%%

\subsection{Data and setup}

The data consist of five selected S\&P500 Sector indices, listed in Table \ref{tab-sp5list}. The sample period spans from January 4, 2006 to December 29, 2017, resulting in $T=3{,}019$ business days. The returns are computed as the log difference of the daily closing price, and demeaned prior to estimation.

%%%%%%%%%%%%%%%%%%%%%%%%%%%%%%%
%%% Stock sector indix list %%%
%%%%%%%%%%%%%%%%%%%%%%%%%%%%%%% 
\begin{table}[b]
\vspace*{5mm}
\centering
\begin{tabular}{cll}

\hline

1	&	INDU	&	Industrials	\\
2	&	MATR	&	Materials	\\
3	&	ENRS	&	Energy	\\
4	&	CONS	&	Consumer Staples \\
5	&	INFT	&	Information Technology \\

\hline

\end{tabular}

\caption{S\&P500 Sector Index.}

\label{tab-sp5list}
\end{table}
%%%%%%%%%%%%%%%%%%%%%%%%%%%%%%

The number of factors is determined as $p=1$ from the conventional factor analysis on the $k=5$ series. The first conventional factor explains 83\% of variations, and the second one only an additional 8\%. A screen-plot shows a significant drop from the first to the second factor. Further, an analysis of predictive performance, which is explained below, confirms that the model with $p=1$ factors performs better than the one with $p=2$.

In theory, there is no ideal method to determine a selection of the first $p$ variables in $\y_t$, which technically induce the factor processes under the triangular identification structure embedded in the factor-loading matrix $\B$. Ideally, those variables are strongly correlated to other series to efficiently extract the common factors. A simple practical approach taken here is to compute a correlation matrix of $\y_t$ and to sum up  each row of the matrix, which represents a degree of correlation with other series. In the dataset, the Industrials sector is the series with the highest degree of correlation, and is selected as the first variable in $\y_t$, as shown in Table \ref{tab-sp5list}.

For the computation, the following priors are assumed: $\mu_i \sim N(-11,1)$, $(\phi_i+1)/2 \sim B(20,1.5)$, $\sigma_i^{-2} \sim G(20, 0.01)$, $(\rho_i+1)/2 \sim B(1, 1)$, $\nu_i \sim G(24, 0.8)I[\nu_i>4]$, $\beta_i \sim \kappa N(\beta_i|0, 10) + (1-\kappa)\delta_0(\beta_i)$, and $\kappa \sim B(2, 2)$, where $B$ and $G$ denotes the beta and gamma distributions, respectively. These priors mainly reflect those assumed in the existing studies~\citep[e.g.,][]{NakajimaOmori12, Nakajima17}. The size of MCMC iteration is 50{,}000 samples after the burn-in period of 5{,}000 samples. To check a convergence of the MCMC draws, the convergence diagnostic of \cite{Geweke92} is computed. In the following analyses, the null that the Markov chain converges is not rejected at 5 percent significance level.

To measure the efficiency of the MCMC algorithm, the relative numerical efficiency measure, or so-called effective sample size is computed~\citep[e.g.,][]{Chib01}. The effective sample size for selected parameters appears to be mostly less than 100, which is comparable to the standard univariate SV models in the existing studies. This figure shows that the developed MCMC algorithm mixes well and yields a practically efficient estimation scheme.

The analysis focuses on the following five FSV specifications:
	\begin{itemize}
	\item {\bf Model S0}: Symmetric $t$-distribution for $\y_t$ and $\f_t$ ($\beta_i\equiv0$).
	\item {\bf Model SY}: Skew $t$-distribution for $\y_t$, symmetric $t$-distribution for $\f_t$ ($\beta_i=0$, for $i>p$), no sparsity on $\beta_i$ for all $i$ (i.e., $\kappa \equiv 1$).
	\item {\bf Model SF}: Skew $t$-distribution for $\f_t$, symmetric $t$-distribution for $\y_t$ ($\beta_i=0$, for $i\le p$), no sparsity on $\beta_i$.
	\item {\bf Model SYF}: Skew $t$-distribution for both $\y_t$ and $\f_t$, no sparsity on $\beta_i$.
	\item {\bf Model SSYF}: Skew $t$-distribution for both $\y_t$ and $\f_t$, with the sparsity prior on $\beta_i$, for all $i$.
	\end{itemize}
A key focus here is on which skewness structure-- for only the idiosyncratic shocks, latent processes, or both of them-- is desirable and how the sparsity skew selection works in forecasting the multivariate returns.

%%%%%%%%%%%%%%%%%%%%%%%%%

\subsection{Posterior estimates}

%%%%%%%%%%%%%%%%%%
%%% Volatility %%%
%%%%%%%%%%%%%%%%%%%%%%%%%%%%%%%%%% 
\begin{figure}[t]

\includegraphics{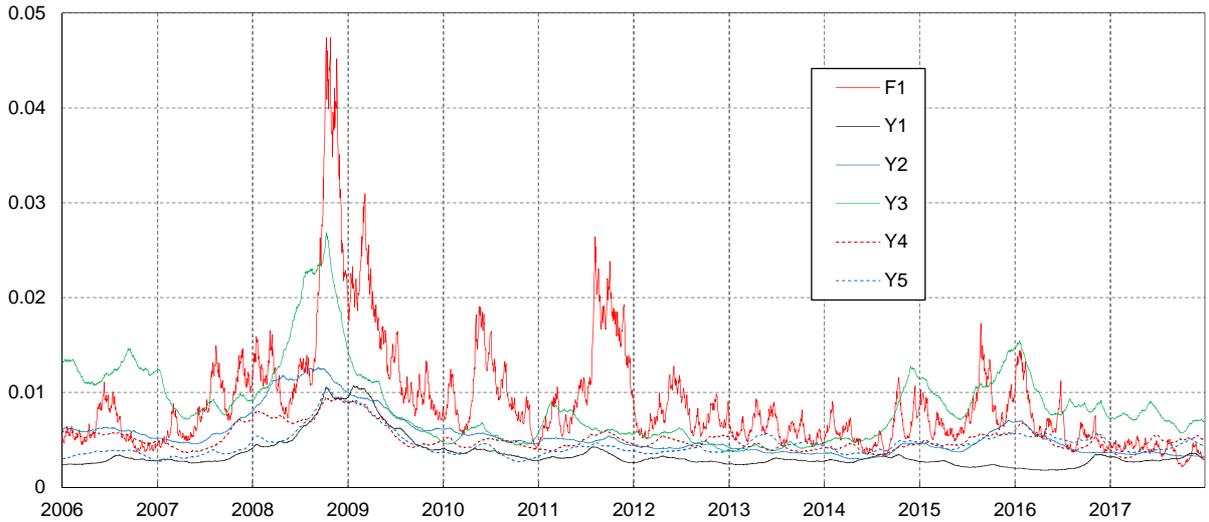}

\caption{Posterior means of the stochastic volatility, $\sigma_{it}=\exp(h_{it}/2)$, from Model {\bf SSYF} for US-sector indices: the factor process (F) and idiosyncratic shocks (Y).}

\label{fig-vol}
\end{figure}
%%%%%%%%%%%%%%%%%%%%%%%%%%%%%%%%%%%%%%%%%%%%%%%%

Prior to the forecasting exercise, posterior estimates based on a fit of the SSYF model are reported. Figure \ref{fig-vol} plots posterior means of the SV process for the factor process and idiosyncratic shocks. The SV of the factor process (F1) exhibits the higher level of volatility and the largest variation among the SV processes, which indicates the variation explained by the latent factor dominates the one by each idiosyncratic shock. The volatilities of five stock indices are highly correlated and their common variation is summarized well by the latent factor. The factor SV increases largely at the time of market turmoil in the global financial crisis around 2009 and the European debt crisis around 2011. The SV of idiosyncratic shock to the Energy sector (Y3) exhibits a different trajectory from  other sectors. In particular, it increases relatively higher than the other series in late 2014 and afterwards. This kind of decoupling in the SV process could reflect an industry-specific business environment. The oil price and other commodity price dropped significantly from late 2014 to 2016, which may have driven up the volatility of stock prices in the energy sector.

%%%%%%%%%%%%%%%%%%
%%% Parameters %%%
%%%%%%%%%%%%%%%%%%%%%%%%%%%%%%%%%% 
\begin{figure}[p]

\hspace{26mm} $\phi_i$ \hspace{47mm} $\sigma_i$  \hspace{47mm} $\rho_i$ \\[-3mm]

\includegraphics[scale = 0.98]{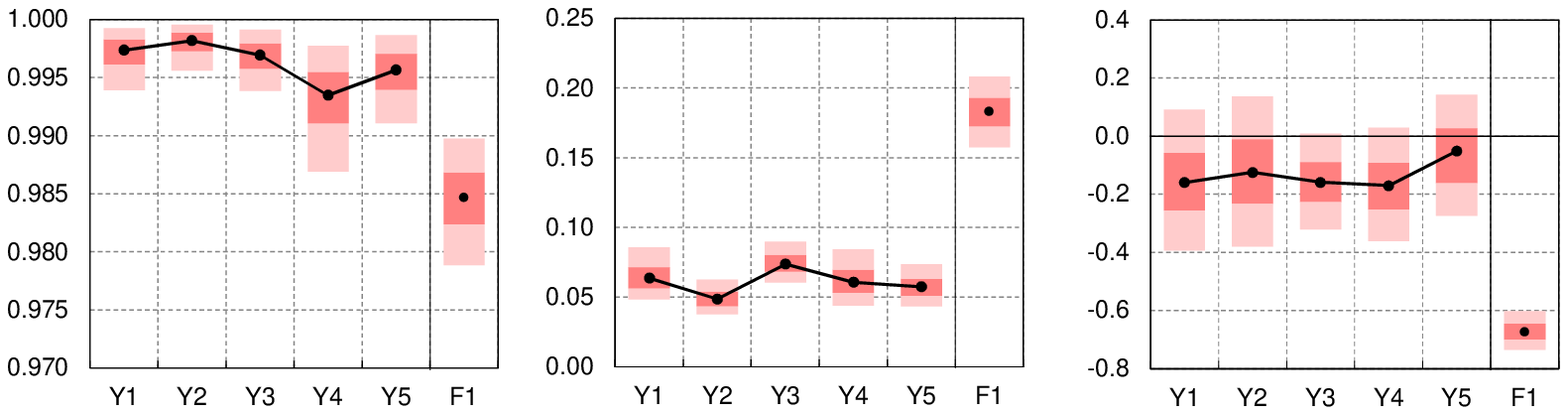}\\[-3mm]

\hspace{26mm} $\mu_i$ \hspace{47mm} $\nu_i$  \hspace{46mm}
$B_{1i}$  \\[-3mm]

\hspace{1mm}
\includegraphics[scale = 0.98]{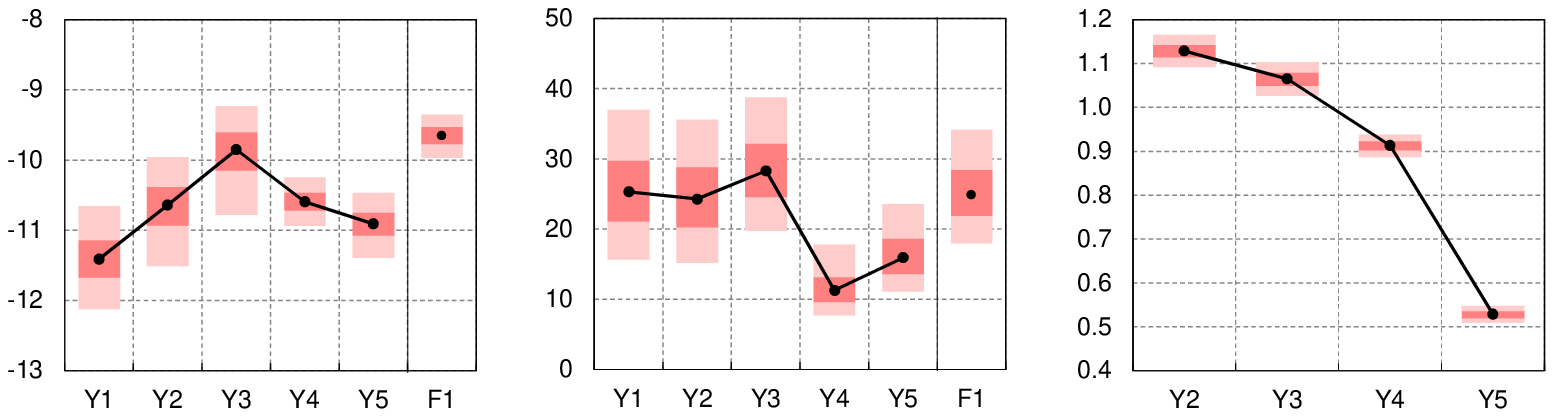}

\caption{Posterior estimates for parameters from Model {\bf SSYF} for the US-sector indices: Posterior medians (dot and solid line), the 50\% (filled area, dark) and 90\% (light) credible intervals. The horizontal axis refers to the idiosyncratic shocks (Y) and factor (F).}

\label{fig-par}
\end{figure}
%%%%%%%%%%%%%%%%%%%%%%%%%%%%%%%%%%%%%%%%%%%%%%%%

%%%%%%%%%%%%%%%%%%%%%%%%%%%%%
%%% Parameters (skewness) %%%
%%%%%%%%%%%%%%%%%%%%%%%%%%%%%%%%%% 
\begin{figure}[p]

\hspace{36mm} $\beta_i$: Model {\bf SYF}
\hspace{26mm} $\beta_i$: Model {\bf SSYF}

\hspace*{23mm}
\includegraphics[scale = 0.98]{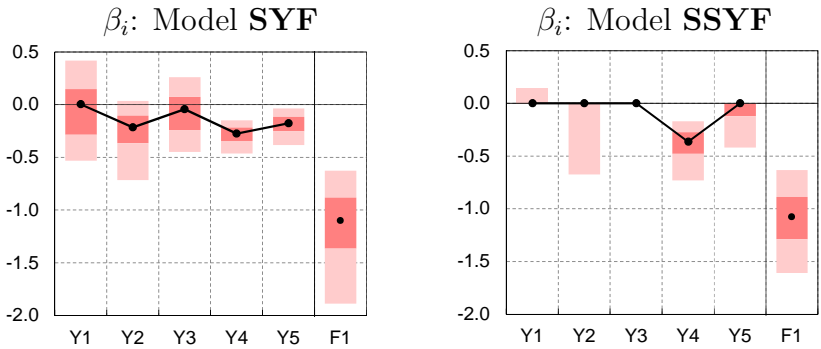}

\caption{Posterior estimates for $\beta_i$ from Models {\bf SYF} (left) and {\bf SSYF} (right) for the US-sector indices: Posterior medians (dot and solid line), the 50\% (filled area, dark) and 90\% (light) credible intervals. The horizontal axis refers to the idiosyncratic shocks (Y) and factor (F).}

\label{fig-pars}
\end{figure}
%%%%%%%%%%%%%%%%%%%%%%%%%%%%%%%%%%%%%%%%%%%%%%%%

Figure \ref{fig-par} shows a result of posterior estimates for selected parameters obtained from the SSYF model, with posterior median exhibited in dots and solid line, the 50\% and 90\% credible intervals in dark and light filled area, respectively.  The posterior median of $\phi_i$ is quite high, around $0.985$ to $0.998$, which suggests a strong persistence in the SV process. The parameter estimates for the factor is notably different from those for the idiosyncratic shocks. The factor process is characterized by a lower value of $\phi_i$ and a higher value of $\sigma_i$ and $\mu_i$ than the idiosyncratic shocks, which implies a larger variation and a higher level of the factor SV process. In addition, the posterior median of $\rho_i$ for the factor is about $-0.4$ with its 90\% credible intervals excluding zero. In contrast, those of all the idiosyncratic shocks include zero. This result suggests that the leverage effect is observed only in the factor process, but not in the idiosyncratic shocks. The degree $\nu_i$ of freedom in the GH skew-$t$ distribution is lower for the Consumer Staples (Y4) and Information Technology (Y5) sectors than the others, which indicates heavier-tails of idiosyncratic shocks in those sectors.

Figure \ref{fig-pars} plots the posterior estimates of the skewness parameter $\beta_i$ from Models SYF and SSYF. For both models, the posterior median of $\beta_i$ for the factor (F1) is about $-1.0$ and the 90\% credible intervals do not include zero. In contrast, $\beta_i$'s for idiosyncratic shocks are mostly less relevant. In Model SYF, the 90\% credible intervals for Y1 to Y3 include zero. The posterior distributions for those sectors are remarkably shrunk towards zero in Model SSYF. While the 90\% credible intervals for Y5 excludes zero in Model SYF, the posterior distribution is partly shrunk in Model SSYF with its posterior median at zero. The posterior probability of $\beta_i=0$ is about 80--95\% for Y1 to Y3 and 70\% for Y5. The Consumer Staples sector (Y4) appears to have relevant skewness in its idiosyncratic shocks with the posterior probability of $\beta_i=0$ only 1\%.

%%%%%%%%%%%%%%%

\subsection{Forecasting performance}

A recursive forecasting exercise is examined to study the model comparison among the competing specifications. The analysis is based on a sequence of updating data and forecasting returns one to five days ahead, in order to illustrate a practical use of the model in the financial market. Specifically, first, the model parameters are estimated for a period, over $t=1\ldots,T_1$, using the posterior computation with the MCMC algorithm. Through the MCMC algorithm, samples of returns $(y_{T_1+1},\ldots,y_{T_1+5})$ are generated by simulating the latent process from one to five days ahead to obtain the posterior predictive density $\pi(y_{T_1+h}|y_1,\ldots,y_{T_1})$ for $h=1,\ldots,5$. We then update the returns for five business days and estimate the model for the returns ($y_1,\ldots,y_{T_1+5}$) using the full MCMC analysis with the computation of posterior predictive density for the next five days, $(y_{T_2+1},\ldots,y_{T_2+5})$, where $T_2=T_1+5$. We repeat this prediction exercise recursively for $T_j=T_{j-1}+5$, $j=2,\ldots,J$, until we obtain forecasts for 500 business days, that is, $J=100$.

We set $T_1=2{,}519$, which corresponds to the first pre-sample period, to January 6, 2016; the forecasting exercise period goes from January 7, 2016 to December 29, 2017. This experiment reflects practical forecasting activity in the financial market, given the historical data over about 10 years at the starting point, recursively forecasting returns in the following week with the updated data. The exercise period is quite long and includes periods of both bull and bear market sentiments.

The forecasting performance of each model is compared based on two diagnostics. The first one is out-of-sample predictive density. Specifically, the log predictive density ratio (LPDR) for Model $i$'s $h$-day ahead forecast is computed as
	\begin{eqnarray*}
	\mathrm{LPDR}_h(M_i) \,=\,\sum_{j=1}^J \log \left\{ \frac{ p_{M_i}(\y_{T_j+h}|\y_{1:T_j})}{p_{M_0}(\y_{T_j+h}|\y_{1:T_j})} \right\},
	\end{eqnarray*}
where $p_M$ denotes the predictive density under model $M$, and $T_1$ denotes the set of time points for the recursive forecasting. This quantity measures how high the likelihood is, when evaluated at the realized return ($\y_{T_j+h}$), that is, how accurate the predictive density is, relative to the base model $M_0$.

Table \ref{tab-dens} reports the result of the LPDR for five models relative to Model S0, showing the LPDR at each horizon $h=1,\ldots,5$ and the sum of all the horizons. Because all the LPDRs are positive, the forecasting performance of the newly introduced skew-$t$ FSV models is superior to the standard FSV model with the symmetric $t$-distribution. Model SF performs better than Model SY, which suggests that the skewness in the factor process is more important for prediction than the ones in the idiosyncratic shocks. Interestingly, Model SYF performs worse than Model SF. This result suggests that the statistically irrelevant skewness parameters for the idiosyncratic shocks, as observed in the left panel of Figure \ref{fig-pars}, worsens the predictive ability of the skew FSV model, unless the shrinkage method is applied. The LPDRs of Model SSYF, which is the highest among the competing models for all the horizons, confirms that the shrinkage method for the skewness parameters effectively selects statistically relevant ones and consequently improves the forecasting performance.

%%%%%%%%%%%%%%%%%%%%%%%%%%
%%% Predictive density %%%
%%%%%%%%%%%%%%%%%%%%%%%%%%%%%%%%%% 
\begin{table}[t]
\centering
\begin{tabular}{lrrrrrr}

\hline
		&	\multicolumn{5}{c}{Horizon ($h$ days)}	&	\\
Model   &	 1     &  2     &  3     &  4	 &  5      & \ Total\  \\
\hline
{\bf SY}     &   4.1  &  28.6  &   90.9 &  26.3 &   54.4  & 204.4 \\
{\bf SF}     & 172.7  & 113.0  &  220.9 & 146.2 &  201.2  & 854.1 \\
{\bf SYF}    &  71.6  &  50.2  &   75.6 &  46.8 &   79.2  & 323.3 \\
{\bf SSYF}   & 195.5  & 204.2  &  231.4 & 146.9 &  216.8  & 994.9 \\

\hline

\end{tabular}

\caption{Cumulative log predictive density ratios ($\mathrm{LPDR}$), relative to Model {\bf S0}, for the US-sector indices.}
\label{tab-dens}
\end{table}
%%%%%%%%%%%%%%%%%%%%%%%%%%%%%%

The second forecasting exercise is based on Value-at-Risk (VaR). The VaR is a standard tool to measure a risk of possible shortfalls in portfolio. We compute the VaR for portfolios that consist of the five stock indices using the forecasts from each competing model, and evaluate the accuracy of the VaR forecast by comparing it with the realized loss of the portfolio return. This exercise focuses on the prediction ability of the skew-$t$ FSV models for a tail risk of the stock returns, which has practical importance in portfolio management.

We consider a standard portfolio allocation strategy based on a mean-variance optimization following \cite{Markowitz59}. The analysis uses standard Bayesian mean-variance optimization. Let $\m_t$ and $\D_t$ denote the forecast mean vector and variance matrix of $\y_t$, respectively. Note that these variables are computed from the posterior predictive distribution obtained in the recursive forecasting, as illustrated in the previous exercise. Define $\bomega_t$ as a vector of portfolio weights, and $\bar{m}$ as a target level of return. We solve the optimization problem with respect to the portfolio weights $\bomega_t$ as it minimizes the forecast variance of the portfolio return among the restricted portfolios whose expectation is equal to $\bar{m}$. We also assume that the amount of money invested into the portfolio is fixed during the sample period, and that the resource can be freely reallocated to arbitrary long or short positions without any transaction cost.

The mathematical form of the optimization problem is minimizing $\bomega_t'\D_t\bomega_t$, subject to $\bomega_t'\m_t = \bar{m}$, and $\bomega_t'\one = 1$. The solution is $\bomega_t^{(\bar{m})} = \K_t (\one' \K_t \q_t \m_t - \m_t' \K_t \q_t \one)$, where $\q_t = (\one \bar{m} - \m_t) / d_t$, and $d_t = (\one' \K_t \one)(\m_t' \K_t \m_t) - (\one' \K_t \m_t)^2$, with $\K_t = \D_t^{-1}$. Note that the portfolio is reallocated every business day. This experiment assumes a practical situation in which investors allocate their resource every day based forecasts updated weekly. The analysis uses a range of daily target returns of $\bar{m}=0.005\%$, $0.01\%$, and $0.02\%$, implying a yearly return of approximately $1.25\%$, $2.5\%$, and $5.0\%$, respectively. In addition, we consider a target-free portfolio that is implemented by minimizing the optimization problem only with $\bomega_t'\one = 1$.

For a diagnostic of model comparison in this VaR forecast exercise, we employ the test statistic of \cite{Kupiec95}. Let $n$ denote the number of VaR violations, and $N$ the total number of forecasting days. Note that 
the expected number of violations for $\alpha$ quantile is $\alpha N$. Under the null hypothesis that the expected ratio of violations is equal to $\alpha$, the likelihood ratio statistic is given by
	\begin{eqnarray*}
	2\log \left\{ \left(\frac{n}{N}\right)^n \left(1-\frac{n}{N}\right)^{N-n} \right\} - 2\log\left\{ \alpha^n (1 - \alpha)^{N-n} \right\}.
	\end{eqnarray*}
As this statistic is asymptotically distributed as $\chi^2(1)$, we use it to test whether the VaR forecast is accurate~\citep[see][]{Kupiec95}. The analysis examines the likelihood ratio test for $\alpha=0.5\%$, $1\%$, and $5\%$ levels.

%%%%%%%%%%%%%%%%
%%% VaR test %%%
%%%%%%%%%%%%%%%%%%%%%%%%%%%%%%%%%% 
\begin{table}[t]
\centering
\begin{tabular}{lrrrrr}

\hline
	  &  &	\multicolumn{3}{l}{Target return ($\bar{m}$)}	&	Target \\
Model &	\multicolumn{1}{c}{$\alpha$} &	0.005\%	&	0.01\%	&	0.02\%	& -free \\
\hline

{\bf S0}   & 0.5\% &  5  \ss  &  4  \ss &  6   \ss &  6\aa   \\
   		   & 1\%   &  9  \ss  &  8  \ss &  7   \ss &  8  \ss \\
           & 5\%   &  31 \ss  &  29 \ss &  29  \ss &  30 \ss \\
{\bf SY}   & 0.5\% &   7\aa   &   7\aa  &   8\aa   &  5  \ss \\
           & 1\%   &  11\aa   &  10\aa  &   8  \ss &  8  \ss \\
           & 5\%   &  26 \ss  &  27 \ss &  31  \ss &  29 \ss \\
{\bf SF}   & 0.5\% &  5  \ss  &  5  \ss &   7\aa   &  3  \ss \\
           & 1\%   &  6  \ss  &  7  \ss &  10\aa   &  8  \ss \\
           & 5\%   &  30 \ss  &  29 \ss &  30  \ss &  27 \ss \\
{\bf SYF}  & 0.5\% &  4  \ss  &  5  \ss &  6\aa    &  5  \ss \\
           & 1\%   &  5  \ss  &  6  \ss &  7   \ss &  8  \ss \\
           & 5\%   &  26 \ss  &  26 \ss &  25  \ss &  26 \ss \\
{\bf SSYF} & 0.5\% &  4  \ss  &  3  \ss &  2   \ss &  3  \ss \\
           & 1\%   &  9  \ss  &  7  \ss &  6   \ss &  7  \ss \\
           & 5\%   &  27 \ss  &  28 \ss &  29  \ss &  26 \ss \\

\hline

\end{tabular}

\caption{VaR results: the number of violations for the US-sector indices. The value with an asterisk ``\aa'' shows that the $p$-value of the likelihood ratio test is less than 10\%. The null hypothesis of the test is that the expected ratio of violations is equal to $\alpha$.}
\label{tab-test}
\end{table}
%%%%%%%%%%%%%%%%%%%%%%%%%%%%%%

Table \ref{tab-test} reports the number of VaR violations and results of the likelihood ratio test. The value with the asterisk (``\aa'') shows that the null hypothesis is rejected at a 10\% significance level. For Model S0, the target-free portfolio does not pass the test. For Models SY, SF, and SYF, some of the target-return portfolios exhibit a rejection of the test. In particular, the number of violations in Model SY indicates its optimistic VaR forecast, possibly due to the lack of skewness parameter in the factor process. Finally, Model SSYF provides no portfolio that is rejected in the selected range of portfolio allocation rules and VaR levels. This confirms that the proposed skew selection strategy works remarkably well for forecasting VaR in portfolio risk management.

%%%%%%%%%%%%%%%

\section{A higher-dimensional study: US individual stocks}

%%%%%%%%%%%%%%%%%%%%%%%%%%%%%%%%
%%% Stock index country list %%%
%%%%%%%%%%%%%%%%%%%%%%%%%%%%%%%% 
\begin{table}[t]
\centering
\begin{tabular}{llll|llll}

\hline

No.	&	Ticker	&	Name	&	Sector	&	No.	&	Ticker	&	Name	&	Sector	\\

\hline

1	&	MS	&	Morgan Stanley	&	FI	&	11 &	TGT	&	Target        	&	CD	\\
2	&	BA	&	Boeing        	&	ID	&	12 &	KO	&	Coca-Cola     	&	CD		\\
3	&	STI 	&	Sun Trust Banks    	&	FI	&	13 &	GIS	&	General Mills 	&	CS		\\
4	&	WFC	&	Wells Fargo   	&	FI	&	14 &	K	&	Kellogg       	&	CS 		\\
5	&	BLK	&	BlackRock     	&	FI	&	15 &	T	&	AT\&T	      		&	TL \\
6	&	MET	&	Met Life      	&	FI	&	16 &	AMT	&	American Tower	&	RE			\\
7	&	AXP	&	American Express	&	FI	&	17 &	COG	&	Cabot Oil \& Gas	&	EN				\\
8	&	L	&	Loews			&	FI	&	18 &	EIX	&	Edison International	&	UT		\\
9	&	IBM	&	IBM		&	IT		&	19 &	FE	&	FirstEnergy			&	UT		\\
10	&	DIS	&	Walt Disney	&	CS	&	20 &	AEE	&	Ameren				&	UT		\\

\hline

\end{tabular}

\caption{List of US individual stocks. The sector categories: Financials (FI), Industrials (ID), Information Technology (IT), Consumer Discretionary (CD), Consumer Staples (CS), Telecommunication Services (TL), Real Estate (RE), Energy (EN), and Utilities (UT).}
\label{tab-listw}
\end{table}
%%%%%%%%%%%%%%%%%%%%%%%%%%%%%%

The second empirical analysis uses daily returns of 20 individual stocks to illustrate a high-dimensional application of the skew FSV models. The stocks are randomly selected from the compositions of S\&P500 index, and listed in Table \ref{tab-listw}. The selected set of securities covers major industrial sectors. The sample period is $T=751$ business days from January 3, 2014 to December 30, 2016. As in the previous analysis, the returns are computed as the log difference of the daily closing price.

We examine the number of factors $p=1$ and $2$ from the conventional factor analysis. The first and second conventional factors explain more than 50\% of variations in total. In addition, an additional factor does not appear to improve the return performance in the following forecasting exercise. The Morgan Stanley and Boeing are selected for the first two series of $\y_t$ for the $p=2$ factors. These two countries from the Financials and Industrials sectors are considered to parsimoniously summarize common movements among the individual stock series. The model with $p=1$ loads Morgan Stanley for the first factor. Note that a robustness check with randomly selected stocks for the first two series revealed that this selection of the first two series do not significantly affect the main result and its implication in this analysis.

A model comparison is examined based on out-of-sample forecasting performance in the recursive forecasting exercise in the same manner as the previous analysis. The posterior predictive distributions are computed for one- to five-day horizons. The first set of data spans to January 5, 2016 and the forecasting period is from January 5, 2016 to December 30, 2016, to obtain the forecasts over 250 business days. Cumulative returns from the optimized portfolios allocations with the target returns ($\bar{m}$) are compared among the competing models. Model SY is excluded from the model comparison because it performs relatively less well in the previous analysis. For the posterior computation, all the prior settings are the same as the previous analysis.

%%%%%%%%%%%%%%%%%%%%%%%%%%
%%% Cumulative returns %%%
%%%%%%%%%%%%%%%%%%%%%%%%%%%%%%%%%% 
\begin{table}[t]
\centering
\begin{tabular}{lrrrr}

\hline
	  &  &	\multicolumn{3}{c}{Target return ($\bar{m}$)}	\\
Model &	 $p$ &	0.005\%	&	0.01\%	&	0.02\%	\\
\hline

{\bf S0}   & 1 & 0.063 & 0.058 & 0.047 \\
{\bf SF}   & 1 & 0.092 & 0.101 & 0.118 \\
{\bf SYF}  & 1 & 0.120 & 0.122 & 0.125 \\
{\bf SSYF} & 1 & 0.147 & 0.153 & 0.165 \\

\hline

{\bf S0}   & 2 & 0.109 & 0.103 & 0.089 \\
{\bf SF}   & 2 & 0.102 & 0.089 & 0.062 \\
{\bf SYF}  & 2 & 0.114 & 0.116 & 0.120 \\
{\bf SSYF} & 2 & 0.121 & 0.130 & 0.147 \\

\hline

\end{tabular}

\caption{Cumulative returns over 250 business days for portfolios obtained from competing models, with the number of factors $p=1$ and $2$.}
\label{tab-cumw}
\end{table}
%%%%%%%%%%%%%%%%%%%%%%%%%%%%%%

Table \ref{tab-cumw} reports the cumulative returns over the forecasting period for portfolios with three target returns. Model SSYF provides the uniformly highest cumulative returns both for $p=1$ and $p=2$ among the competing models for each target return. The return from the $p=1$ SSYF model performs slightly better than the $p=2$ model. This result confirms that the skew FSV model with parsimonious skew selection structure improves the forecasting performance. Interestingly, Model SF is outperformed by Model S0 for $p=2$, while Model SYF outperforms Model S0, which indicates that the skewness for idiosyncratic shocks is partly relevant to describe dynamics of the high-dimensional multivariate returns.

%%%%%%%%%%%%%%%%%%
%%% Parameters %%%
%%%%%%%%%%%%%%%%%%%%%%%%%%%%%%%%%% 
\begin{figure}[p]

\hspace{38mm} $\phi_i$ \hspace{75mm} $\sigma_i$ \\[-3mm]

\hspace{-2mm}
\includegraphics{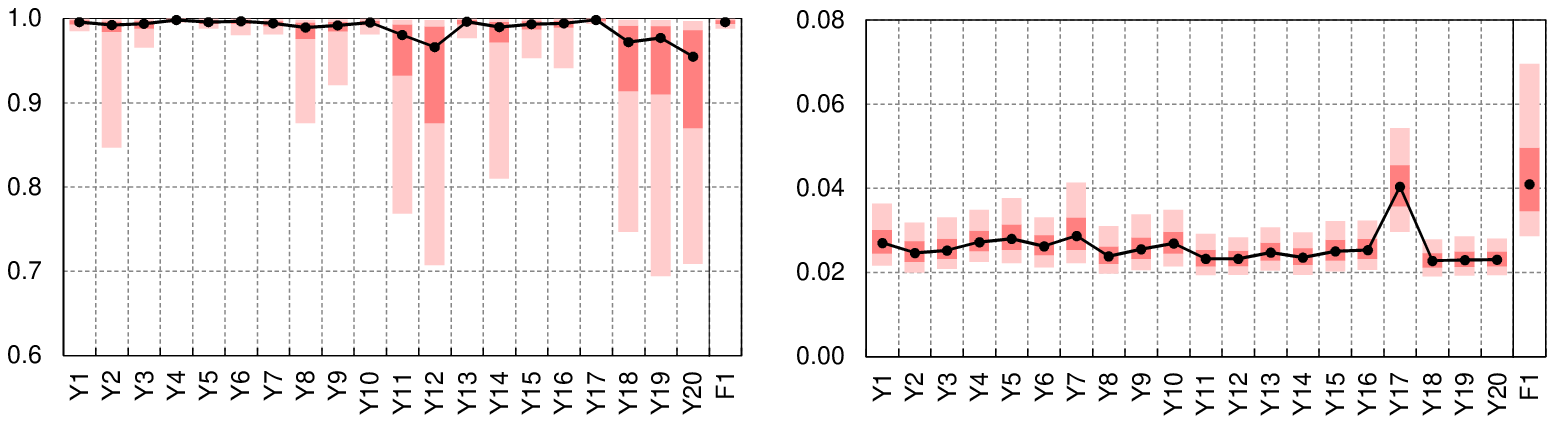}\\[-3mm]

\hspace{-2mm}
\hspace{38mm} $\rho_i$ \hspace{75mm} $\mu_i$ \\[-3mm]

\hspace{-3mm}
\includegraphics{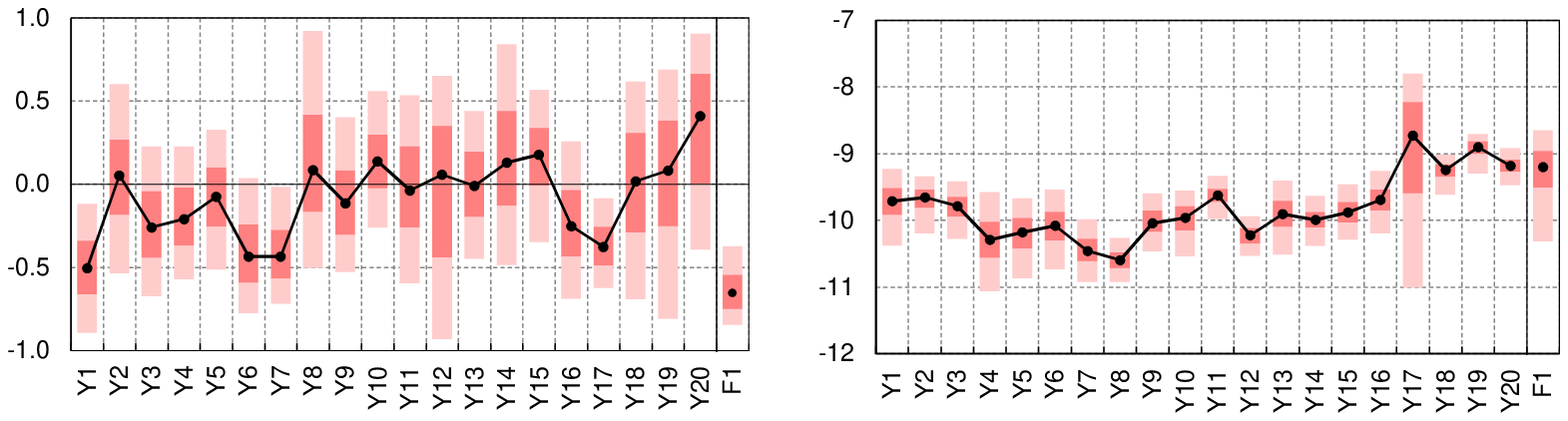}\\[-3mm]

\hspace{38mm} $\nu_i$ \hspace{75mm} $\beta_i$ \\[-3mm]

\includegraphics{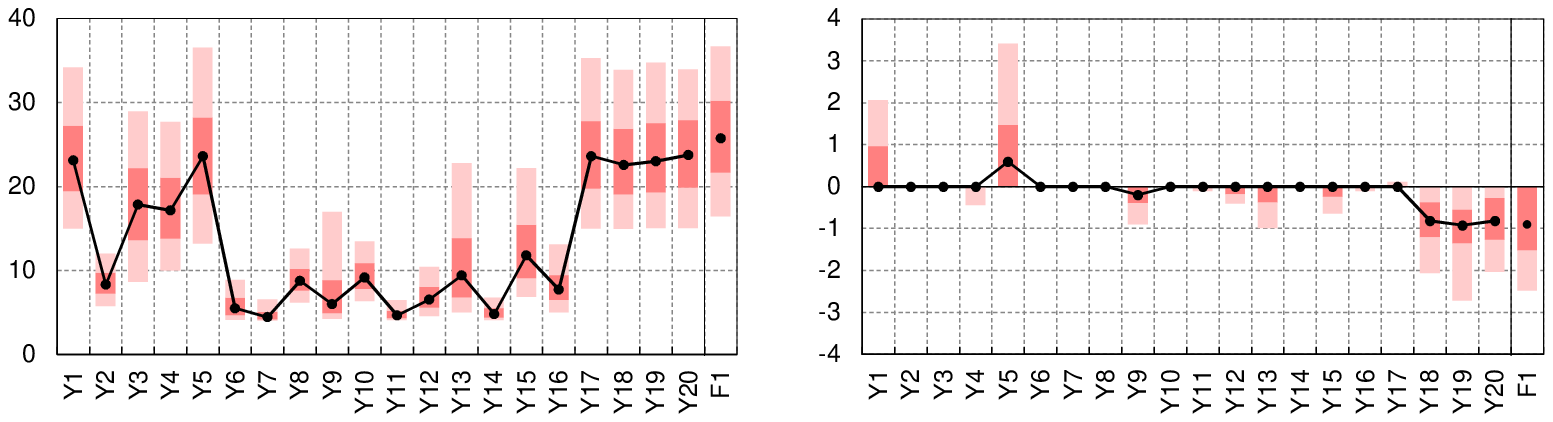}\\[-3mm]

\hspace{76mm} $B_{1i}$ \\[-3mm]

\hspace{40mm}\includegraphics{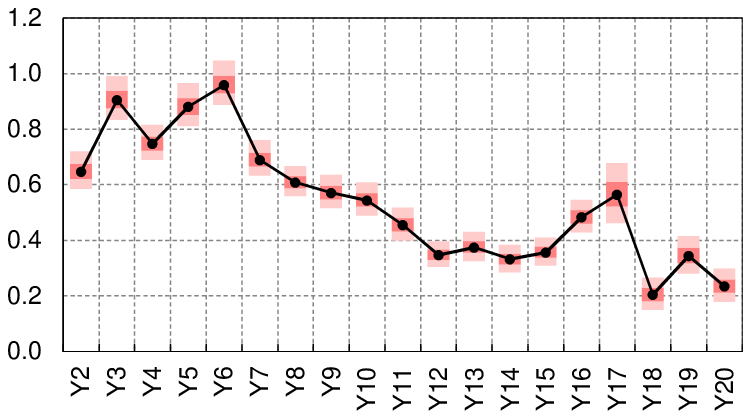}\\[-3mm]

\caption{Posterior estimates for parameters from Model {\bf SSYF} with $p=1$ for US individual stock returns: Posterior medians (dot and solid line), the 50\% (filled area, dark) and 90\% (light) credible intervals. The horizontal axis refers to the idiosyncratic shocks (Y) and factor (F).}

\label{fig-parw}
\end{figure}
%%%%%%%%%%%%%%%%%%%%%%%%%%%%%%%%%%%%%%%%%%%%%%%%

Figure \ref{fig-parw} shows posterior estimates of selected parameters from Model SSYF with $p=1$. The posterior estimates of $\phi_i$ indicates quite high persistence for the SV processes of both the factors and idiosyncratic shocks, while some of the idiosyncratic shocks exhibit relatively wider credible intervals. For the parameter $\rho_i$, the posterior median of the factor is about $-0.7$, suggesting that the leverage effect exists as a common behavior among the individual stocks. Additionally, several idiosyncratic shocks (Y1, Y7, Y17) exhibit relevant leverage effect. The posterior estimates of the degree $\nu_i$ of the freedom are quite low for several stocks, with some of the posterior medians being below 10. This estimate indicates that the price movement of individual stocks is relatively more volatile and the return distribution is more heavily-tailed than the sector indices examined in the previous analysis.

The estimates of $\beta_i$ show considerable shrinkage for most of the idiosyncratic shocks. While some of them exhibit that the posterior medians or credible intervals depart from zero, 16 of 20 idiosyncratic shocks exhibit a posterior median at zero. For Y18 to Y20, the skewness is relatively more relevant than the other series with the 50\% credible intervals excluding zero. The result in the forecasting performance partly reflects this relevant skewness for idiosyncratic shocks. The $\beta_i$ for the factor is around $-1.0$, which indicates the importance of the skewness in the factor process.

%%%%%%%%%%%%%%%%%%%%%%%%%

\section{Concluding remarks}

This paper develops the FSV model with GH skew-$t$ distribution. The Bayesian sparsity prior is embedded to the skewness parameters, which provides a parsimonious skew structure for the FSV model. Two empirical analyses using daily stock returns of US-sector indices and individual stocks show that the advantage of the skew selection structure that effectively shrinks irrelevant skewness parameters and consequently improves forecasting performance.

Possible extensions can be explored in several directions. The factor loadings can be time-varying and the factor process can be more complicated dynamics, such as VAR and VARMA processes. From another perspective, it is of interest to model a time-varying skewness that take non-zero values when relevant and zero otherwise using time-varying sparsity techniques~\citep[e.g.][]{NakajimaWest12jbes,HuberKastnerFeldkircher19}.

%%%%%%%%%%%%%%%%%%%%%%%%%

\begin{spacing}{1.4}

\bibliographystyle{chicago}
\bibliography{svskf}

\begin{thebibliography}{}

\bibitem[\protect\citeauthoryear{Aas and Haff}{Aas and Haff}{2006}]{AasHaff06}
Aas, K. and I.~H. Haff (2006).
\newblock The generalized hyperbolic skew {S}tudent's t-distribution.
\newblock {\em Journal of Financial Econometrics\/}~{\em 4}, 275--309.

\bibitem[\protect\citeauthoryear{Abanto-Valle, Lachos, and Dey}{Abanto-Valle
  et~al.}{2015}]{AbantoValleLachosDey15}
Abanto-Valle, C.~A., V.~H. Lachos, and D.~K. Dey (2015).
\newblock Bayesian estimation of a skew-student-t stochastic volatility model.
\newblock {\em Methodology and Computing in Applied Probability\/}~{\em 17},
  721--738.

\bibitem[\protect\citeauthoryear{Aguilar, Prado, Huerta, and West}{Aguilar
  et~al.}{1999}]{Aguilar1999}
Aguilar, O., R.~Prado, G.~Huerta, and M.~West (1999).
\newblock Bayesian inference on latent structure in time series (with
  discussion).
\newblock In J.~M. Bernardo, J.~O. Berger, A.~P. Dawid, and A.~F.~M. Smith
  (Eds.), {\em Bayesian Statistics 6}, pp.\  3--26. Oxford: Oxford University
  Press.

\bibitem[\protect\citeauthoryear{Aguilar and West}{Aguilar and
  West}{2000}]{AguilarWest00}
Aguilar, O. and M.~West (2000).
\newblock Bayesian dynamic factor models and portfolio allocation.
\newblock {\em Journal of Business and Economic Statistics\/}~{\em 18},
  338--357.

\bibitem[\protect\citeauthoryear{Azzalini and Capitanio}{Azzalini and
  Capitanio}{2003}]{AzzaliniCapitanio03}
Azzalini, A. and A.~Capitanio (2003).
\newblock Distributions generated by pertubation of symmetry with emphasis on a
  multivariate skew $t$ distribution.
\newblock {\em Journal of the Royal Statistical Society B\/}~{\em 65},
  367--389.

\bibitem[\protect\citeauthoryear{Barndorff-Nielsen and
  Shephard}{Barndorff-Nielsen and Shephard}{2001}]{Barndorff-NielsenShephard01}
Barndorff-Nielsen, O.~E. and N.~Shephard (2001).
\newblock {\em Normal modified stable rocesses}.
\newblock University of Oxford.

\bibitem[\protect\citeauthoryear{Black}{Black}{1976}]{Black76}
Black, F. (1976).
\newblock Studies of stock market volatility changes.
\newblock In {\em Proceedings of the American Statistical Association, Business
  and Economic Statistics Section}, pp.\  177--181.

\bibitem[\protect\citeauthoryear{Carvalho, Lopes, and Aguilar}{Carvalho
  et~al.}{2011}]{CarvalhoLopesAguilar11}
Carvalho, C.~M., H.~F. Lopes, and O.~Aguilar (2011).
\newblock Dynamic stock selection strategies: {A} structured factor model
  framework (with discussion).
\newblock In J.~M. Bernardo, M.~J. Bayarri, J.~O. Berger, A.~P. Dawid,
  D.~Heckerman, A.~F.~M. Smith, and M.~West (Eds.), {\em Bayesian Statistics
  9}, pp.\  69--90. Oxford University Press.

\bibitem[\protect\citeauthoryear{Chib}{Chib}{2001}]{Chib01}
Chib, S. (2001).
\newblock Markov chain {M}onte {C}arlo methods: computation and inference.
\newblock In J.~J. Heckman and E.~Leamer (Eds.), {\em Handbook of
  Econometrics}, Volume~5, pp.\  3569--3649. Amsterdam: North-Holland.

\bibitem[\protect\citeauthoryear{Chib, Nardari, and Shephard}{Chib
  et~al.}{2006}]{ChibNardariShephard06}
Chib, S., F.~Nardari, and N.~Shephard (2006).
\newblock Analysis of high dimensional multivariate stochastic volatility
  models.
\newblock {\em Journal of Econometrics\/}~{\em 134}, 341--371.

\bibitem[\protect\citeauthoryear{Chib, Omori, and Asai}{Chib
  et~al.}{2009}]{ChibOmoriAsai09}
Chib, S., Y.~Omori, and M.~Asai (2009).
\newblock Multivariate stochastic volatility.
\newblock In T.~G. Andersen, R.~A. Davis, J.-P. Kreiss, and T.~Mikosch (Eds.),
  {\em Handbook of Financial Time Series}, pp.\  365--400. Springer-Verlag
  Berlin Heidelberg.

\bibitem[\protect\citeauthoryear{Clyde and George}{Clyde and
  George}{2004}]{ClydeGeorge04}
Clyde, M. and E.~I. George (2004).
\newblock Model uncertainty.
\newblock {\em Statistical Science\/}~{\em 19}, 81--94.

\bibitem[\protect\citeauthoryear{Eberlein, Keller, and Prause}{Eberlein
  et~al.}{1998}]{EberleinKellerPrause98}
Eberlein, E., U.~Keller, and K.~Prause (1998).
\newblock New insighs into smile, mispricing and value at risk: the hyperbolic
  model.
\newblock {\em Journal of Business\/}~{\em 71}, 371--405.

\bibitem[\protect\citeauthoryear{Fern\'andez and Steel}{Fern\'andez and
  Steel}{1998}]{FernandezSteel98}
Fern\'andez, C. and M.~F.~J. Steel (1998).
\newblock On {B}ayesian modeling of fat tails and skewness.
\newblock {\em Journal of the American Statistical Association\/}~{\em 93},
  359--371.

\bibitem[\protect\citeauthoryear{Fr\"uwirth-Schnatter and
  Lopes}{Fr\"uwirth-Schnatter and Lopes}{2018}]{FruhwirthLopes18}
Fr\"uwirth-Schnatter, S. and H.~F. Lopes (2018).
\newblock Parsimonious {B}ayesian factor analysis when the number of factors is
  unknown.
\newblock Working Paper.

\bibitem[\protect\citeauthoryear{George and McCulloch}{George and
  McCulloch}{1993}]{GeorgeMcCulloch93}
George, E.~I. and R.~E. McCulloch (1993).
\newblock Variable selection via {G}ibbs sampling.
\newblock {\em Journal of the American Statistical Association\/}~{\em 88},
  881--889.

\bibitem[\protect\citeauthoryear{George and McCulloch}{George and
  McCulloch}{1997}]{GeorgeMcCulloch97}
George, E.~I. and R.~E. McCulloch (1997).
\newblock Approaches for {B}ayesian variable selection.
\newblock {\em Statistica Sinica\/}~{\em 7}, 339--373.

\bibitem[\protect\citeauthoryear{Geweke}{Geweke}{1992}]{Geweke92}
Geweke, J. (1992).
\newblock Evaluating the accuracy of sampling-based approaches to the
  calculation of posterior moments.
\newblock In J.~M. Bernardo, J.~O. Berger, A.~P. Dawid, and A.~F.~M. Smith
  (Eds.), {\em Bayesian Statistics}, Volume~4, pp.\  169--188. New York: Oxford
  University Press.

\bibitem[\protect\citeauthoryear{Geweke and Zhou}{Geweke and
  Zhou}{1996}]{GewekeZhou96}
Geweke, J.~F. and G.~Zhou (1996).
\newblock Measuring the pricing error of the arbitrage pricing theory.
\newblock {\em Review of Financial Studies\/}~{\em 9}, 557--587.

\bibitem[\protect\citeauthoryear{Han}{Han}{2005}]{Han05}
Han, Y. (2005).
\newblock Asset allocation with a high dimensional latent factor stochastic
  volatility model.
\newblock {\em Review of Financial Studies\/}~{\em 19}, 237--271.

\bibitem[\protect\citeauthoryear{Hansen}{Hansen}{1994}]{Hansen94}
Hansen, B.~E. (1994).
\newblock Autoregressive conditional density estimation.
\newblock {\em International Economic Review\/}~{\em 35}, 705--730.

\bibitem[\protect\citeauthoryear{Huber, Kastner, and Feldkircher}{Huber
  et~al.}{2019}]{HuberKastnerFeldkircher19}
Huber, F., G.~Kastner, and M.~Feldkircher (2019).
\newblock Should {I} stay or should {I} go? {A} latent threshold approach to
  large-scale mixture innovation models.
\newblock \textit{Journal of Applied Econometrics}, forthcoming.

\bibitem[\protect\citeauthoryear{Ishihara and Omori}{Ishihara and
  Omori}{2017}]{IshiharaOmori17}
Ishihara, T. and Y.~Omori (2017).
\newblock Portfolio optimization using dynamic factor and stochastic
  volatility: evidence on fat-tailed error and leverage.
\newblock {\em Japanese Economic Review\/}~{\em 68}, 63--94.

\bibitem[\protect\citeauthoryear{Jones and Faddy}{Jones and
  Faddy}{2003}]{JonesFaddy03}
Jones, M.~C. and M.~J. Faddy (2003).
\newblock A skew extension of the $t$-distribution, with application.
\newblock {\em Journal of Royal Statistical Society, Series B\/}~{\em 65},
  159--174.

\bibitem[\protect\citeauthoryear{Kastner}{Kastner}{2018}]{Kastner18}
Kastner, G. (2018).
\newblock Sparse {B}ayesian time-varying covariance estimation in many
  dimensions.
\newblock \textit{Journal of Econometrics}, forthcoming.

\bibitem[\protect\citeauthoryear{Kobayashi}{Kobayashi}{2016}]{Kobayashi16}
Kobayashi, G. (2016).
\newblock Skew exponential power stochastic volatility model for analysis of
  skewness, non-normal tails, quantiles and expectiles.
\newblock {\em Computational Statistics\/}~{\em 31}, 49--88.

\bibitem[\protect\citeauthoryear{Kupiec}{Kupiec}{1995}]{Kupiec95}
Kupiec, P. (1995).
\newblock Techniques for verifying the accuracy of risk measurement models.
\newblock {\em Journal of Derivatives\/}~{\em 2}, 173--184.

\bibitem[\protect\citeauthoryear{Langrock, Michelot, Sohn, and Kneib}{Langrock
  et~al.}{2015}]{LangrockMichelotSohnKneib15}
Langrock, R., T.~Michelot, A.~Sohn, and T.~Kneib (2015).
\newblock Semiparametric stochastic volatility modelling using penalized
  splines.
\newblock {\em Computational Statistics\/}~{\em 30}, 517--537.

\bibitem[\protect\citeauthoryear{Li and Scharth}{Li and
  Scharth}{2018}]{LiScharth18}
Li, M. and M.~Scharth (2018).
\newblock Leverage, asymmetry and heavy tails in the high-dimensional factor
  stochastic volatility model.
\newblock Working Paper Series, 2018-49, Economics Discipline Group, UTS
  Business School, University of Technology, Sydney.

\bibitem[\protect\citeauthoryear{Loddo, Ni, and Sun}{Loddo
  et~al.}{2011}]{LoddoNiSun11}
Loddo, A., S.~Ni, and D.~Sun (2011).
\newblock Selection of multivariate stochastic volatility models via {B}ayesian
  stochastic search.
\newblock {\em Journal of Business \& Economic Statistics\/}~{\em 29},
  342--355.

\bibitem[\protect\citeauthoryear{Lopes and Carvalho}{Lopes and
  Carvalho}{2007}]{LopesCarvalho07}
Lopes, H.~F. and C.~M. Carvalho (2007).
\newblock Factor stochastic volatility with time varying loadings and {M}arkov
  switching regimes.
\newblock {\em Journal of Statistical Planning and Inference\/}~{\em 137},
  3082--3091.

\bibitem[\protect\citeauthoryear{Lopes and West}{Lopes and
  West}{2004}]{LopesWest04}
Lopes, H.~F. and M.~West (2004).
\newblock Bayesian model assessment in factor analysis.
\newblock {\em Statistica Sinica\/}~{\em 14}, 41--67.

\bibitem[\protect\citeauthoryear{Markowitz}{Markowitz}{1959}]{Markowitz59}
Markowitz, H. (1959).
\newblock {\em Portfolio Selection: Efficient Diversification of Investments}.
\newblock New York: John Wiley and Sons.

\bibitem[\protect\citeauthoryear{McNeil, Frey, and Embrechts}{McNeil
  et~al.}{2005}]{McNeil05}
McNeil, A.~J., R.~Frey, and P.~Embrechts (2005).
\newblock {\em Quantitative Risk Management: Concepts, Techniques, and Tools}.
\newblock New Jersey: Princeton University Press.

\bibitem[\protect\citeauthoryear{Nakajima}{Nakajima}{2017}]{Nakajima17}
Nakajima, J. (2017).
\newblock Bayesian analysis of multivariate stochastic volatility with skew
  distribution.
\newblock {\em Econometric Reviews\/}~{\em 36}, 546--562.

\bibitem[\protect\citeauthoryear{Nakajima and Omori}{Nakajima and
  Omori}{2012}]{NakajimaOmori12}
Nakajima, J. and Y.~Omori (2012).
\newblock Stochastic volatility model with leverage and asymmetrically
  heavy-tailed error using {GH} skew student's $t$-distribution.
\newblock {\em Computational Statistics and Data Analysis\/}~{\em 56},
  3690--3704.

\bibitem[\protect\citeauthoryear{Nakajima and West}{Nakajima and
  West}{2013a}]{NakajimaWest12jbes}
Nakajima, J. and M.~West (2013a).
\newblock Bayesian analysis of latent threshold dynamic models.
\newblock {\em Journal of Business and Economic Statistics\/}~{\em 31},
  151--164.

\bibitem[\protect\citeauthoryear{Nakajima and West}{Nakajima and
  West}{2013b}]{NakajimaWest12jfe}
Nakajima, J. and M.~West (2013b).
\newblock Dynamic factor volatility modeling: {A} {B}ayesian latent threshold
  approach.
\newblock {\em Journal of Financial Econometrics\/}~{\em 11}, 116--153.

\bibitem[\protect\citeauthoryear{Nelson}{Nelson}{1991}]{Nelson91}
Nelson, D.~B. (1991).
\newblock Conditional heteroskedasticity in asset returns: a new approach.
\newblock {\em Econometrica\/}~{\em 59}, 347--370.

\bibitem[\protect\citeauthoryear{Omori, Chib, Shephard, and Nakajima}{Omori
  et~al.}{2007}]{OmoriChibShephardNakajima07}
Omori, Y., S.~Chib, N.~Shephard, and J.~Nakajima (2007).
\newblock Stochastic volatility with leverage: fast likelihood inference.
\newblock {\em Journal of Econometrics\/}~{\em 140}, 425--449.

\bibitem[\protect\citeauthoryear{Omori and Watanabe}{Omori and
  Watanabe}{2008}]{OmoriWatanabe08}
Omori, Y. and T.~Watanabe (2008).
\newblock Block sampler and posterior mode estimation for asymmetric stochastic
  volatility models.
\newblock {\em Computational Statistics and Data Analysis\/}~{\em 52},
  2892--2910.

\bibitem[\protect\citeauthoryear{Pitt and Shephard}{Pitt and
  Shephard}{1999}]{PittShephard99b}
Pitt, M. and N.~Shephard (1999).
\newblock Time varying covariances: {A} factor stochastic volatility approach
  (with discussion).
\newblock In J.~M. Bernardo, J.~O. Berger, A.~P. Dawid, and A.~F.~M. Smith
  (Eds.), {\em Bayesian Statistics VI}, pp.\  547--570. Oxford University
  Press.

\bibitem[\protect\citeauthoryear{Prado and West}{Prado and
  West}{2010}]{PradoWest10}
Prado, R. and M.~West (2010).
\newblock {\em Time Series Modeling, Computation, and Inference}.
\newblock New York: Chapman \& Hall/CRC.

\bibitem[\protect\citeauthoryear{Prause}{Prause}{1999}]{Prause99}
Prause, K. (1999).
\newblock The {G}eneralized {H}yperbolic models: Estimation, financial
  derivatives and risk measurement.
\newblock PhD dissertation, University of Freiburg.

\bibitem[\protect\citeauthoryear{Takahashi, Watanabe, and Omori}{Takahashi
  et~al.}{2016}]{TakahashiWatanabeOmori16}
Takahashi, M., T.~Watanabe, and Y.~Omori (2016).
\newblock Volatility and quantile forecasts by realized stochastic volatility
  models with generalized hyperbolic distribution.
\newblock {\em International Journal of Forecasting\/}~{\em 32}, 437--457.

\bibitem[\protect\citeauthoryear{West}{West}{2003}]{West2003}
West, M. (2003).
\newblock Bayesian factor regression models in the ``large p, small n''
  paradigm.
\newblock In J.~Bernardo, M.~Bayarri, J.~Berger, A.~David, D.~Heckerman,
  A.~Smith, and M.~West (Eds.), {\em Bayesian Statistics 7}, pp.\  723--732.
  Oxford.

\bibitem[\protect\citeauthoryear{Yu}{Yu}{2005}]{Yu05}
Yu, J. (2005).
\newblock On leverage in a stochastic volatility model.
\newblock {\em Journal of Econometrics\/}~{\em 127}, 165--178.

\bibitem[\protect\citeauthoryear{Zhou, Nakajima, and West}{Zhou
  et~al.}{2014}]{ZhouNakajimaWest14}
Zhou, X., J.~Nakajima, and M.~West (2014).
\newblock Bayesian forecasting and portfolio decisions using dynamic dependent
  sparse factor models.
\newblock {\em International Journal of Forecasting\/}~{\em 30}, 963--980.

\end{thebibliography}

\end{spacing}

\end{document}